\documentclass[twocolumn]{aastex63}

\usepackage{color}
\usepackage{xspace}
\usepackage{graphicx}
\usepackage{amsmath}
\usepackage{amssymb}
\usepackage{yfonts}
\usepackage{xcolor}
\hypersetup{pdfauthor={Name}}

\definecolor{darkorange}{rgb}{1.0, 0.55, 0.0}

\def\grb{GRB\,210812A\xspace}
\def\swift{{\it Swift}}
\def\fermi{{\it Fermi}}
\def\Ep{E_{\rm peak}}

\newcommand{\Msun}{M_{\odot}}

\graphicspath{{./}{./}} 

\shorttitle{Lensed GRB\,210812A}
\shortauthors{Veres et al.}

\begin{document}
\sloppy

\title{Fermi-GBM observations of GRB\,210812A: signatures of a million solar mass gravitational lens
}

\correspondingauthor{P. Veres}
\email{peter.veres@uah.edu}

\def\aj{AJ}
\def\actaa{Acta Astron.}
\def\araa{ARA\&A}
\def\apj{ApJ}
\def\apjl{ApJ}
\def\apjs{ApJS}
\def\ao{Appl.~Opt.}
\def\apss{Ap\&SS}
\def\aap{A\&A}
\def\aapr{A\&A~Rev.}
\def\aaps{A\&AS}
\def\azh{AZh}
\def\baas{BAAS}
\def\bac{Bull. astr. Inst. Czechosl.}
\def\caa{Chinese Astron. Astrophys.}
\def\cjaa{Chinese J. Astron. Astrophys.}
\def\icarus{Icarus}
\def\jcap{J. Cosmology Astropart. Phys.}
\def\jrasc{JRASC}
\def\mnras{MNRAS}
\def\memras{MmRAS}
\def\na{New A}
\def\nar{New A Rev.}
\def\pasa{PASA}
\def\pra{Phys.~Rev.~A}
\def\prb{Phys.~Rev.~B}
\def\prc{Phys.~Rev.~C}
\def\prd{Phys.~Rev.~D}
\def\pre{Phys.~Rev.~E}
\def\prl{Phys.~Rev.~Lett.}
\def\pasp{PASP}
\def\pasj{PASJ}
\def\qjras{QJRAS}
\def\rmxaa{Rev. Mexicana Astron. Astrofis.}
\def\skytel{S\&T}
\def\solphys{Sol.~Phys.}
\def\sovast{Soviet~Ast.}
\def\ssr{Space~Sci.~Rev.}
\def\zap{ZAp}
\def\nat{Nature}
\def\iaucirc{IAU~Circ.}
\def\aplett{Astrophys.~Lett.}
\def\apspr{Astrophys.~Space~Phys.~Res.}
\def\bain{Bull.~Astron.~Inst.~Netherlands}
\def\fcp{Fund.~Cosmic~Phys.}
\def\gca{Geochim.~Cosmochim.~Acta}
\def\grl{Geophys.~Res.~Lett.}
\def\jcp{J.~Chem.~Phys.}
\def\jgr{J.~Geophys.~Res.}
\def\jqsrt{J.~Quant.~Spec.~Radiat.~Transf.}
\def\memsai{Mem.~Soc.~Astron.~Italiana}
\def\nphysa{Nucl.~Phys.~A}
\def\physrep{Phys.~Rep.}
\def\physscr{Phys.~Scr}
\def\planss{Planet.~Space~Sci.}
\def\procspie{Proc.~SPIE}
\let\astap=\aap
\let\apjlett=\apjl
\let\apjsupp=\apjs
\let\applopt=\ao
%


\def\ve{\varepsilon}
\def\astar{A_\star}
\def\et3{\eta_3}
\def\th1{\theta_{-1}}
\def\r07{r_{0,7}}
\def\x05{x_{0.5}}
\def\ve{\varepsilon}
\def\muh{\hat{\mu}}
\def\cm{\hbox{~cm}}
\def\km{\hbox{~km}}
\def\pc{\hbox{~pc}}
\def\kpc{\hbox{~kpc}}
\def\Mpc{\hbox{~Mpc}}
\def\Gpc{\hbox{~Gpc}}
\def\s{\hbox{~s}}
\def\gev{\hbox{~GeV}}
\def\uJy{~\mu\hbox{Jy}}
\def\Jy{\hbox{~Jy}}
\def\Hz{\hbox{~Hz}}
\def\TeV{\hbox{~TeV}}
\def\GeV{\hbox{~GeV}}
\def\MeV{\hbox{~MeV}}
\def\kev{\hbox{~keV}}
\def\keV{\hbox{~keV}}
\def\eV{\hbox{~eV}}
\def\G{\hbox{~G}}
\def\erg{\hbox{~erg}}
\def\s{{\hbox{~s}}}
\def\ms{{\hbox{~ms}}}
\def\para{\parallel}
\def\Fl{\mathcal{F}}
\def\Ep{E_{\rm peak}}
\def\Epk{E$_{\rm peak}$}
\def\Eiso{E$_{\rm iso}$}
\def\Liso{L$_{\rm iso}$}
\def\T90{T$_{\rm 90}$}
\def\tv{t_{\rm var}}
\def\Lt{L_{\hbox{erg s}^{-1}}}
\def\Lobs{{1.6 \times 10^{47}~ \hbox{erg s}^{-1}}}
\def\LobsOLD{{1.6 \times 10^{48}~ \hbox{erg s}^{-1}}} 
\def\r0{1.2 \times 10^{6}~ \hbox{cm}}
\def\es{\hbox{~erg s}^{-1}}
\def\fr#1#2{{{#1} \over {#2}}}

\newcommand{\NU}{\affiliation{Center for Interdisciplinary Exploration and Research in Astrophysics (CIERA) and Department of Physics and Astronomy, Northwestern University, 1800 Sherman Ave, Evanston, IL 60201, USA}}
\newcommand{\GSFC}{\affiliation{NASA Goddard Space Flight Center, University of Maryland, Baltimore County, Greenbelt, MD 20771, USA}}
\newcommand{\CfA}{\affiliation{Center for Astrophysics\:$|$\:Harvard \& Smithsonian, 60 Garden St. Cambridge, MA 02138, USA}}
\newcommand{\Einstein}{\altaffiliation{NASA Einstein Fellow}}
\newcommand{\NASA}{\altaffiliation{NASA Postdoctoral Fellow}}
\newcommand{\UAH}{\affiliation{Center for Space Plasma and Aeronomic Research, University of Alabama in Huntsville, 320 Sparkman Drive, Huntsville, AL 35899, USA}}
\newcommand{\UNAM}{\affiliation{Instituto de Astronom{\'\i}a, Universidad Nacional Aut\'onoma de M\'exico, Apartado Postal 70-264, 04510 M\'exico, CDMX, Mexico}}
\newcommand{\USRA}{\affiliation{Science and Technology Institute, Universities Space Research Association, Huntsville, AL 35805, USA}}
\newcommand{\Arizona}{\affiliation{University of Arizona, Steward Observatory, 933 N. Cherry Avenue, Tucson, AZ 85721, USA}}
\newcommand{\Bath}{\affiliation{Department of Physics, University of Bath, Claverton Down, Bath, BA2 7AY, UK}}
\newcommand{\OU}{\affiliation{Astrophysical Institute, Department of Physics and Astronomy, 251B Clippinger Lab, Ohio University, Athens, OH 45701, USA}}
\newcommand{\Adler}{\affiliation{The Adler Planetarium, Chicago, IL 60605, USA}}
\newcommand{\GeminiN}{\affiliation{Gemini Observatory/NSF's NOIRLab, 670 N. A'ohoku Place, Hilo, HI, 96720, USA}}
\newcommand{\UMD}{\affiliation{Joint Space-Science Institute, University of Maryland, College Park, MD 20742, USA}}
\newcommand{\GWU}{\affiliation{Department of Physics, The George Washington University, Washington, DC 20052, USA}}
\newcommand{\Leicester}{\affiliation{School of Physics and Astronomy, University of Leicester, University Road, Leicester, LE1 7RH, UK}}
\newcommand{\Marin}{\affiliation{College of Marin, 120 Kent Avenue, Kentfield 94904 CA, USA}}
\newcommand{\UVI}{\affiliation{University of the Virgin Islands, \#2 Brewers bay road, Charlotte Amalie, 00802 USVI, USA}}
\newcommand{\Radboud}{\affiliation{Department of Astrophysics/IMAPP, Radboud University, 6525 AJ Nijmegen, The Netherlands}}
\newcommand{\Warwick}{\affiliation{Department of Physics, University of Warwick, Coventry, CV4 7AL, UK}}
\newcommand{\Birmingham}{\affiliation{Birmingham Institute for Gravitational Wave Astronomy and School of Physics and Astronomy, University of Birmingham, Birmingham B15 2TT, UK}}
\newcommand{\Edinburgh}{\affiliation{Institute for Astronomy, University of Edinburgh, Royal Observatory, Blackford Hill, EH9 3HJ, UK}}
\newcommand{\Caltech}{\affiliation{Cahill Center for Astrophysics, California Institute of Technology, 1200 E. California Blvd. Pasadena, CA 91125, USA}}
\newcommand{\LJMU}{\affiliation{Astrophysics Research Institute, Liverpool John Moores University, 146 Brownlow Hill, Liverpool L3 5RF, UK}}
\newcommand{\CCA}{\affiliation{Center for Computational Astrophysics, Flatiron Institute, 162 W. 5th Avenue, New York, NY 10011, USA}}
\newcommand{\Columbia}{\affiliation{Department of Physics and Columbia Astrophysics Laboratory, Columbia University, New York, NY 10027, USA}}
\newcommand{\CRESST}{\affiliation{Center for Research and Exploration in Space Science and Technology (CRESST) and NASA Goddard Space Flight Center, Greenbelt, MD 20771, USA}}
\newcommand{\Maryland}{\affiliation{Department of Physics, University of Maryland, Baltimore County, 1000 Hilltop Circle, Baltimore, MD 21250, USA}} 
%
%
\author[0000-0002-2149-9846]{P.~Veres}
\UAH
\author{N.~Bhat}
\UAH
\author[0000-0002-0173-6453]{N.~Fraija}
\UNAM
%
%
\author[0000-0001-8058-9684]{S.~Lesage}
\UAH
%
%
%
%
%
%
%
%
%
%
%
%
%
%
%
%
%
%
%
%
%
%
%
%

\begin{abstract}
Observing gravitationally lensed objects in the time domain is difficult, and well-observed time-varying sources are rare.   Lensed gamma-ray bursts (GRBs) offer improved timing precision to this class of objects complementing observations of quasars and supernovae.  The rate of lensed GRBs is highly uncertain, approximately 1 in 1000. The  Gamma-ray Burst Monitor (GBM) onboard the Fermi Gamma-ray Space Telescope has observed more than 3000 GRBs making it an ideal instrument to uncover lensed bursts. Here we present observations of GRB 210812A showing two emission episodes, separated by 33.3 s, and with flux ratio of about 4.5. An exhaustive temporal and spectral analysis shows that the two emission episodes have the same pulse and spectral shape, which poses challenges to GRB models.  We report multiple lines of evidence for a gravitational lens origin. In particular, modeling the lightcurve using nested sampling we uncover strong evidence in favor of the lensing scenario. Assuming a point mass lens, the mass of the lensing object is about 1 million solar masses. High-resolution radio imaging is needed for future lens candidates to derive tighter constraints.
\end{abstract}

\keywords{gamma-ray burst --- gamma-ray transient source}

\section{Introduction}\label{sec:intro}
Strong gravitational lensing is a tool that serendipitously enhances our observing capabilities and offers new opportunities to study the Universe \citep[see e.g.][]{Oguri19lens}. Gamma-ray bursts (GRBs) are energetic transient sources at cosmological distances, involving relativistic jets from stellar-mass black hole (BH) central engines.  GRBs last from a fraction of a second to about 1000 s and typically show non-thermal spectra \citep[see e.g.][for reviews]{Kumar2015,Beloborodov+17review}. Given that the distance scale of GRBs spans a wide range (up to redshift $z\lesssim9$, \citet{Cucchiara2011}), a fraction of GRBs will show  the imprints of strong gravitational lensing. 

Strong gravitational lensing produces multiple images of the same source.  The images differ in their intensity, but importantly their spectral shapes will be the same.  Similarly, in time-varying sources, the temporal profile will be the same but shifted in time for different images \citep{Schneider+96lensbook}. The temporal and spectral invariance is the main defining feature of gravitational lensing. Lensed images are separated by up to arcseconds which are clearly below the resolution of current gamma-ray detectors. Gamma-ray instruments, however, have excellent time resolution, and temporal structures can be recorded with unparalleled accuracy \citep{Meegan2009}. 

In one incarnation of the lensing scenario \citep{1964MNRAS.128..307R,Rodney+21SNlens} applied to GRBs, also called macrolensing, a single event triggers the {same} instrument twice, and can be separated anywhere from a few hours to decades. The two triggers will have similar lightcurve shapes and spectra. The delay between the events scales linearly with the lens mass. 
{Lens candidates for this scenario have masses in the $10^{8}-10^{12}M_\odot$ range.
Objects in this mass range include supermassive black holes (up to few $\times 10^{10} M_{\odot}$), galaxies and clusters of galaxies. In practice, however, all traditional strong lensing time delay measurements are from galaxies or clusters of galaxies.
}
GRB lensing rates are highly uncertain but somewhere on the order of 1 in 1000 GRBs should be affected \citep{Mao92lens}. After roughly 10,000 observed GRBs during three decades, no convincing macrolensing candidate GRB pair has been found  \citep{Nemiroff+94lens,Davidson+11lens, Veres+09lens,Hurley+19lens, Ahlgren+20lens}. The negative result is likely a combination of two effects: First, GRB detectors typically in low Earth orbit will miss a sizeable fraction of GRB lens echoes \citep[see, however][for existing and future all-sky instruments]{Hurley+19lens,Hui+21moonbeam}. Second, for weaker GRBs, with pulses close to the noise level, it is difficult to distinguish between the lensing scenario and just two unrelated but similar looking GRBs \citep{Ahlgren+20lens}.

In a different scenario, also called millilensing \citep[][because the expected separation between the images is on the order of milli-arcseconds]{Nemiroff+01millilens}, the gravitational lens signature is imprinted upon the lightcurve of a single trigger. In this case, we have, e.g., two emission episodes with similar lightcurve patterns which can be separated by timescales spanning from a fraction of a second to a few minutes.

Recently, there has been an increase in claims of millilensing events. \citet{Paynter+21lens} presented convincing evidence for lensing in the short duration BATSE GRB 950830. \citet{Mukherjee+21lens} raised some concerns based on the inconsistent flux ratio between the two pulses. 
\citet{Yang+21lens} and \citet{Wang+21lens} independently argued that the likely short GBM GRB 200716C shows milli-lensing signatures. \citet{Kalantari+21lens} reported on a different GBM lensing candidate, the long duration GRB 090717, selected based on the analysis of the auto-correlation function. The claim of \citet{Kalantari+21lens} was challenged by \citet{Mukherjee+21lenslctest} arguing that  the two pulse shapes differ significantly. 
For the above-reported lensing candidates the flux of the first and the second emission episode is either at the same level or their ratio is $\lesssim 1.5$.

We present observations of the long duration \grb and show that it is consistent with a gravitational lensing scenario. It is the first lensing claim with a flux ratio $\gtrsim 3$. We list multiple lines of evidence to support the lensing interpretation.
We perform spectral and temporal analysis using \fermi-GBM data and complement it with additional data from INTEGRAL-SPI/ACS and Swift/BAT.

In Section \ref{sec:obs}, we present the observations, followed by tests for gravitational lensing in Section \ref{sec:test}. We discuss the power of the tests in Section \ref{sec:discussion} and conclude in Section \ref{sec:conclusion}.

\section{Observations}\label{sec:obs}
\begin{figure}
	\centering
	\includegraphics[width=\columnwidth]{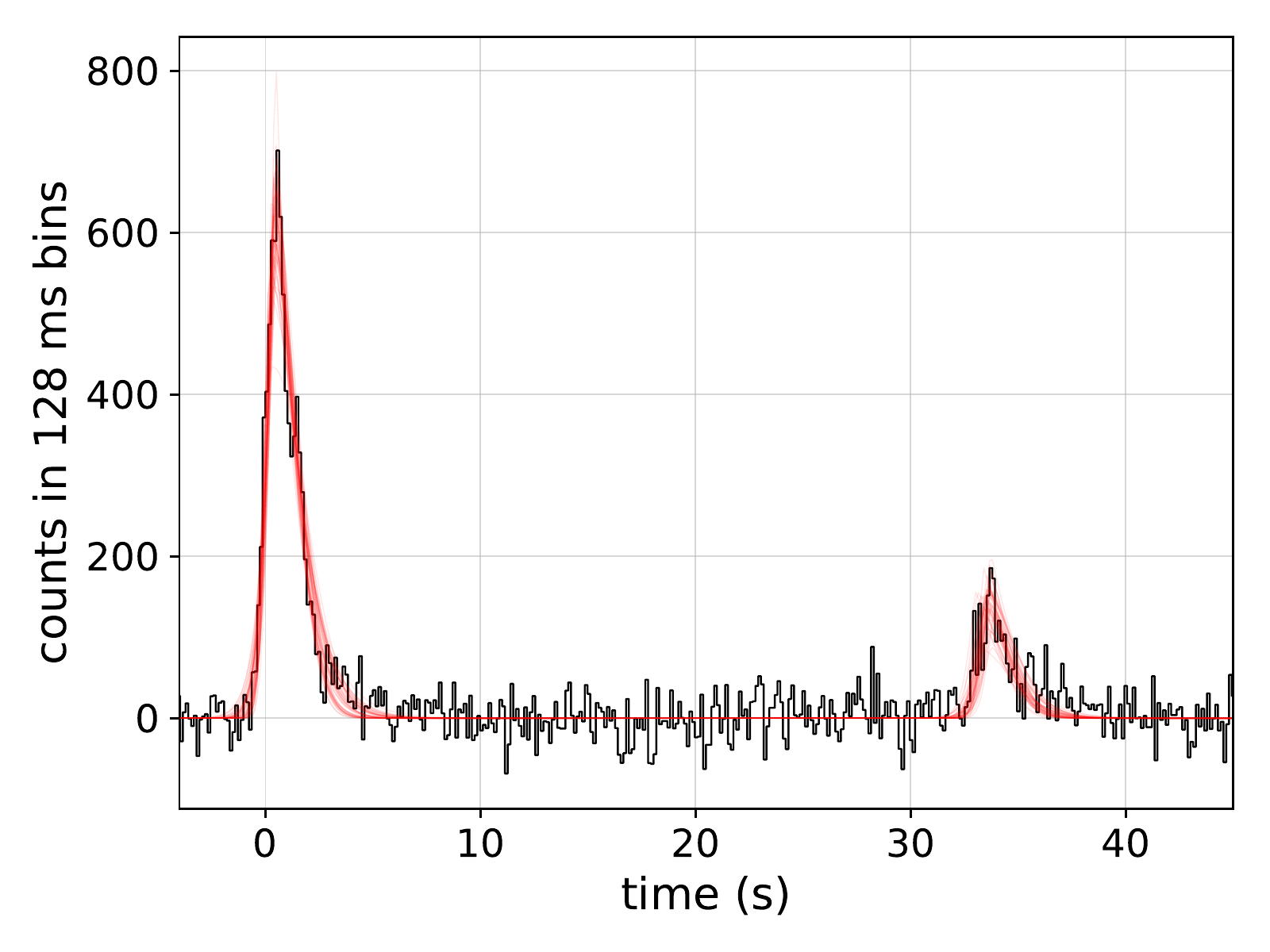}
    \caption{Background subtracted lightcurve of \grb. The data is summed over all GBM detectors with good coverage. The red lines show a subset of MCMC  fits  using the N1 pulse (see Equation \ref{eq:n1} and Section \ref{sec:mcmc}).}
    \label{fig:lcfit}
\end{figure}

\fermi-GBM triggered on \grb (Figure \ref{fig:lcfit}, trigger number 650479626 / 210812699) on August 12$^{\rm th}$, 2021 at 16:47:01.014 UT (T0) \citep{Fermi21lensgcn,Veres21lensgcn}. 
\fermi-GBM consists of 12 NaI (referred to as n0, n1,..., n9, na and nb) and 2 BGO (referred to as b0 and b1) detectors covering the entire unocculted sky in the 8-1000 keV and $\sim$0.1-40 MeV energy range, respectively.
%

As reported by the automatic pipeline, the GRB location is RA$=39.7$, dec$=69.7$ degrees, with an error radius of $1.1$ degrees (statistical only). At the trigger time, this position corresponds to a LAT boresight angle of 149 degrees. The GRB location was behind the spacecraft, meaning most GBM detector normals have a large angle to the source.
Based on previous experience \citep[e.g.][]{Connaughton+16gwgrb}, this type of geometry results in a large number of detectors showing approximately equal count rates compared to $\sim$3 detectors with dominating signal for typical GRBs with small LAT boresight angle.
Upon visual inspection of the NaI detectors' lightcurves in the 50-300 keV range, where GRBs are brightest, we find that all 12 NaI detectors detected the brighter first peak. Moreover, detectors n1 and n6 through nb also detected the fainter, second pulse. Both of the pulses were detected by the 2 BGO detectors.

Detectors n8, na, and nb showed the strongest signals. We used data from these detectors, along with b1, for spectral analysis. For the temporal analysis, we used detectors n1, n6 through nb, b0, and b1. For both spectral and temporal analysis, we used the 128 energy channel time-tagged event (\texttt{TTE}) data. We chose the pre-binned, 8 energy channel (\texttt{ctime}) data to carry out the flux-ratio test. 

The \texttt{targeted search} \citep{Blackburn+15targeted,Goldstein+19targeted} was designed to search for coherent sub-threshold signals (weak signals that did 
not trigger the instrument). As expected, we recovered both pulses of \grb with high significance. The search also provides a location for the burst (RA$=40.5^\circ$, dec$=69.4^\circ$), consistent with the location of the automatic pipeline used for standard GRB analysis. We also find that the locations of the two pulses are consistent, meaning they do indeed belong to the same source.

\subsection{Other observations}

The location of \grb was outside the coded mask of \swift-BAT \citep{Gehrels2004,Barthelmy2005} at the time of the trigger. The first peak of \grb is clearly present in the continuous 4-channel data\footnote{
\url{https://www.swift.ac.uk/archive/reproc/00096363010/bat/rate/sw00096363010brtms.lc.gz}},\footnote{
\url{https://www.swift.ac.uk/archive/reproc/00036688035/bat/rate/sw00036688035brtms.lc.gz}
}, but the second peak is only discernible in the summed lightcurve.

\begin{figure}
	\centering
	\includegraphics[width=\columnwidth]{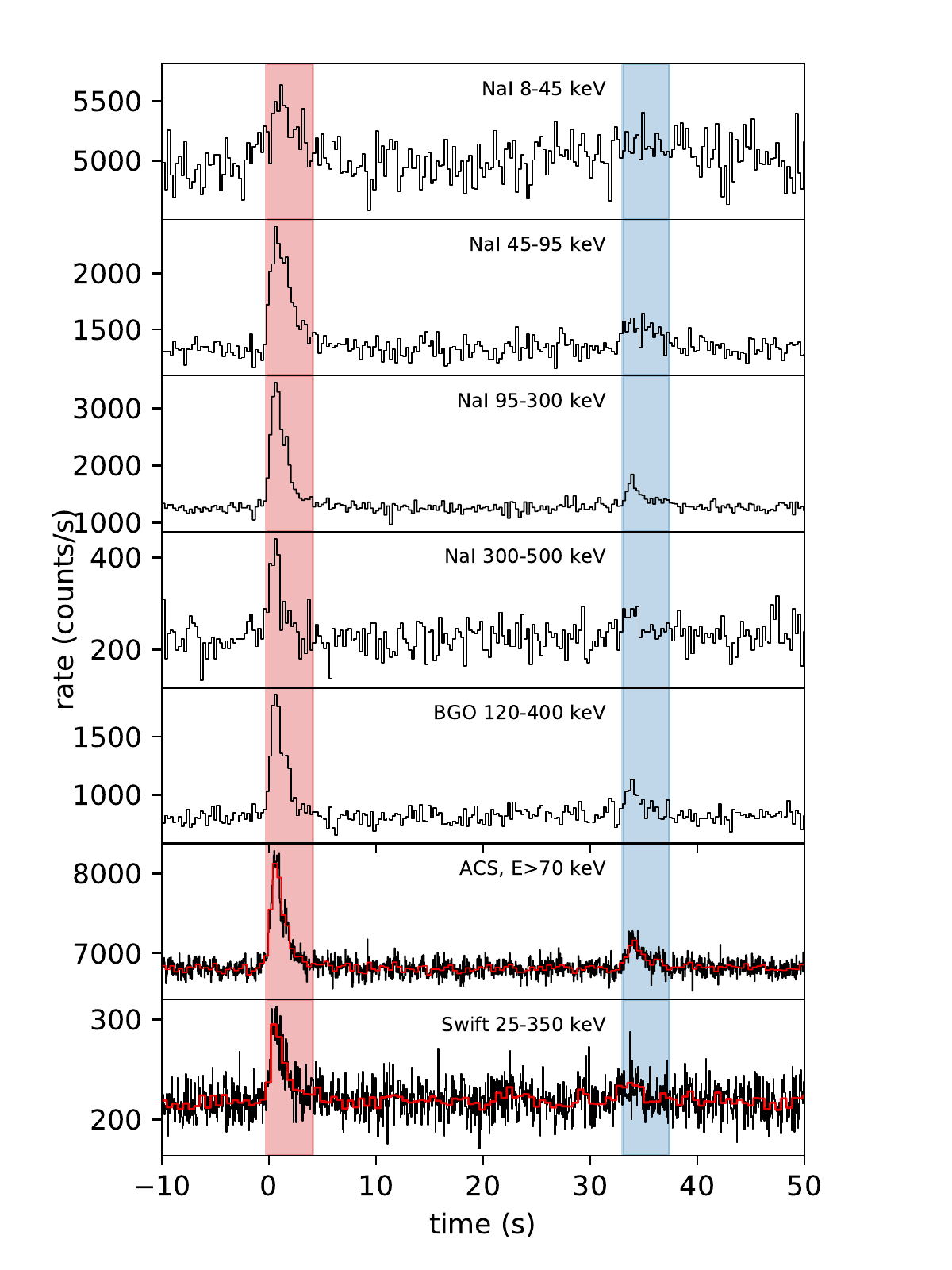}
    \caption{GBM lightcurves in different energy ranges, summed over all detectors with good coverage. ACS and Swift lightcurves are also shown in the bottom two panels with the native resolution (black, 50 ms for ACS and 64 ms for BAT) and binned (red, 0.4 s for ACS and 0.512 s for BAT).}
    \label{fig:multilc}
\end{figure}

INTEGRAL/SPI-ACS \citep{Winkler+03integral,vonKienlin+03acs} detected \grb and clearly shows the two-peak structure.\footnote{\url{http://isdc.unige.ch/~savchenk/spiacs-online/spiacs.pl}} We calculate the time delay between \fermi -GBM and ACS is due to the higher altitude of the INTEGRAL spacecraft to be $dt_{\rm ACS}=-0.396~{\rm ms}$. We applied this correction to the ACS lightcurve in Figure \ref{fig:multilc}.

\begin{figure}
	\centering
		\includegraphics[width=0.49\columnwidth]{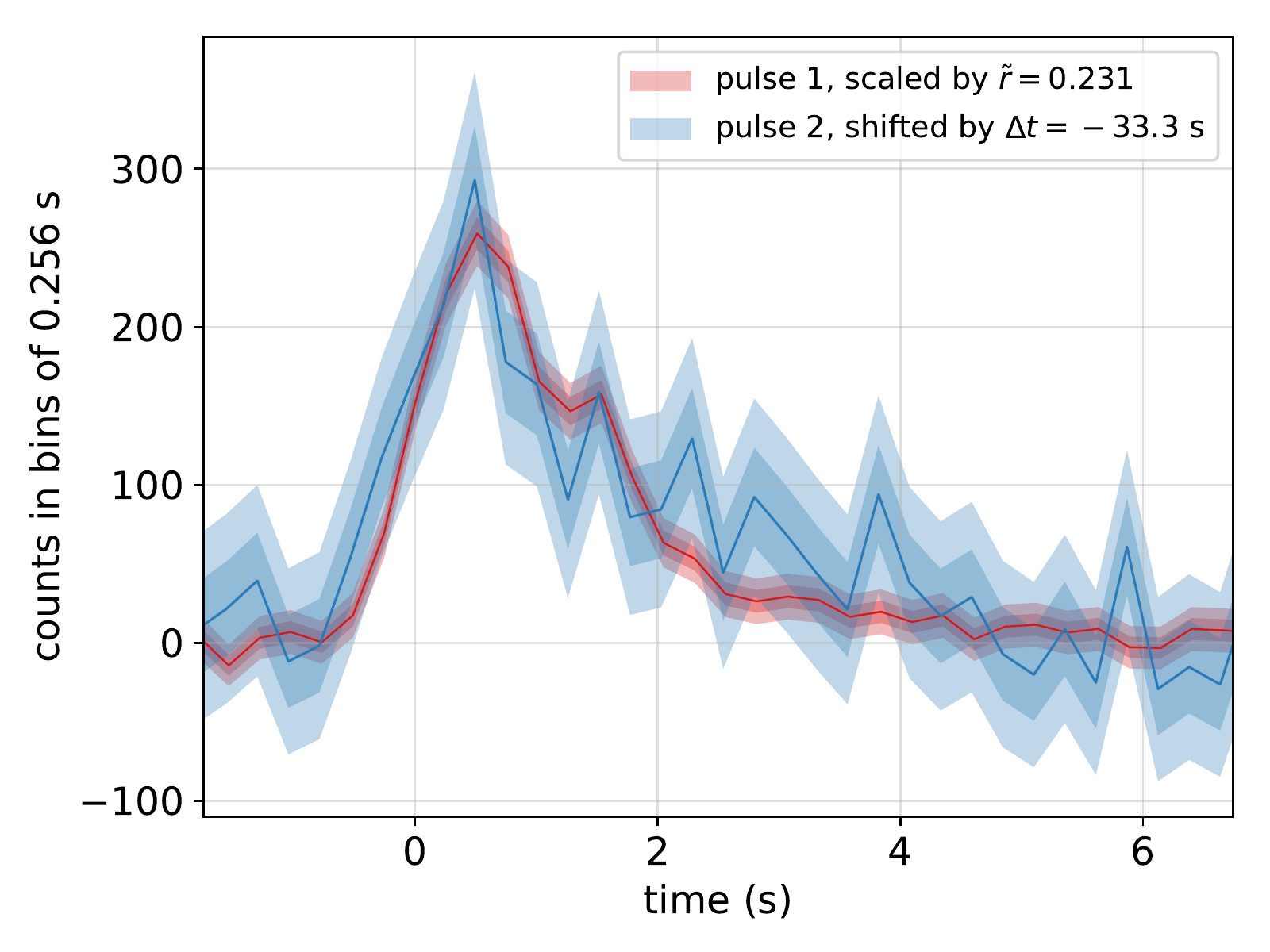}
	\includegraphics[width=0.49\columnwidth]{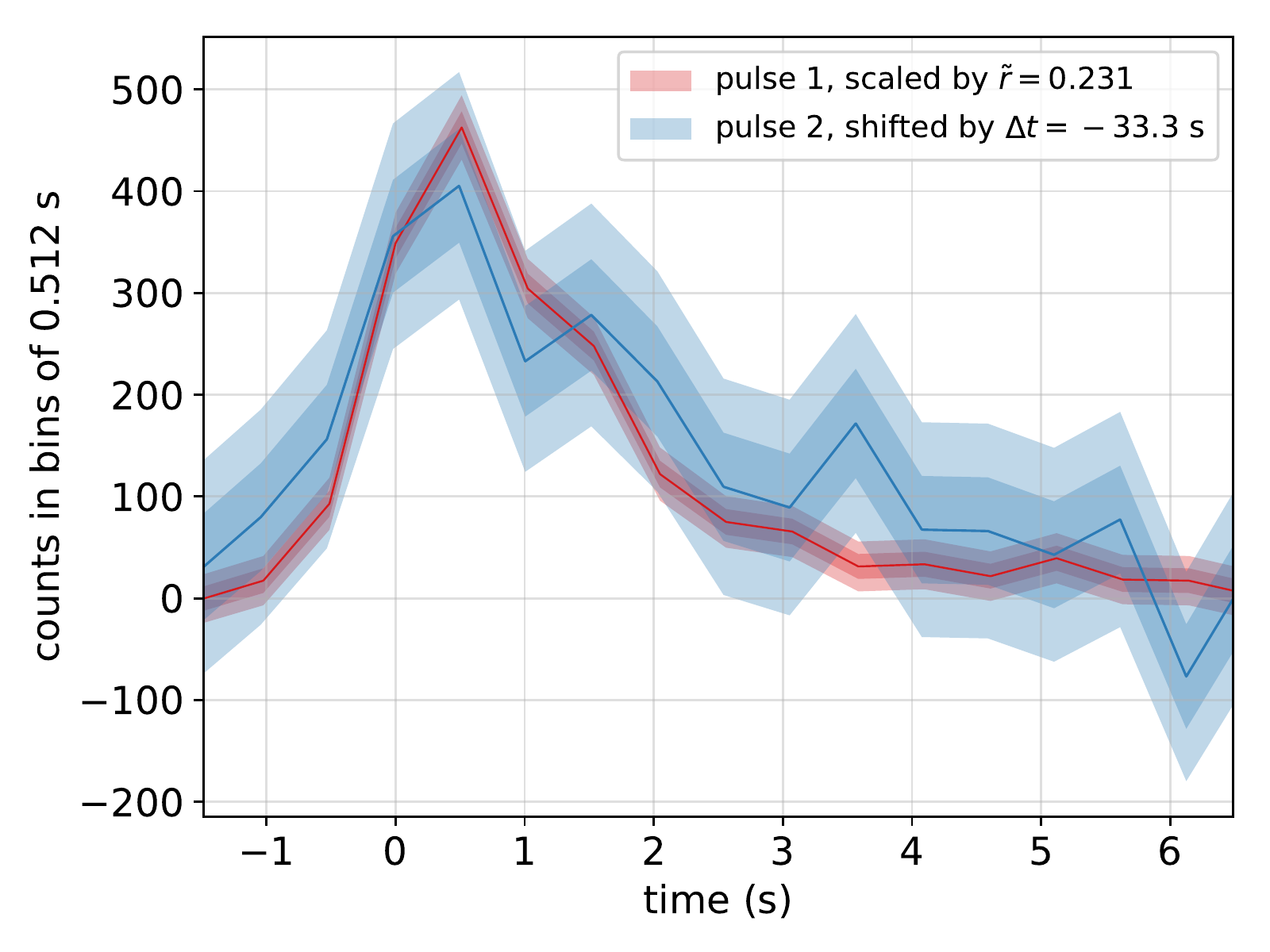}
	\includegraphics[width=0.49\columnwidth]{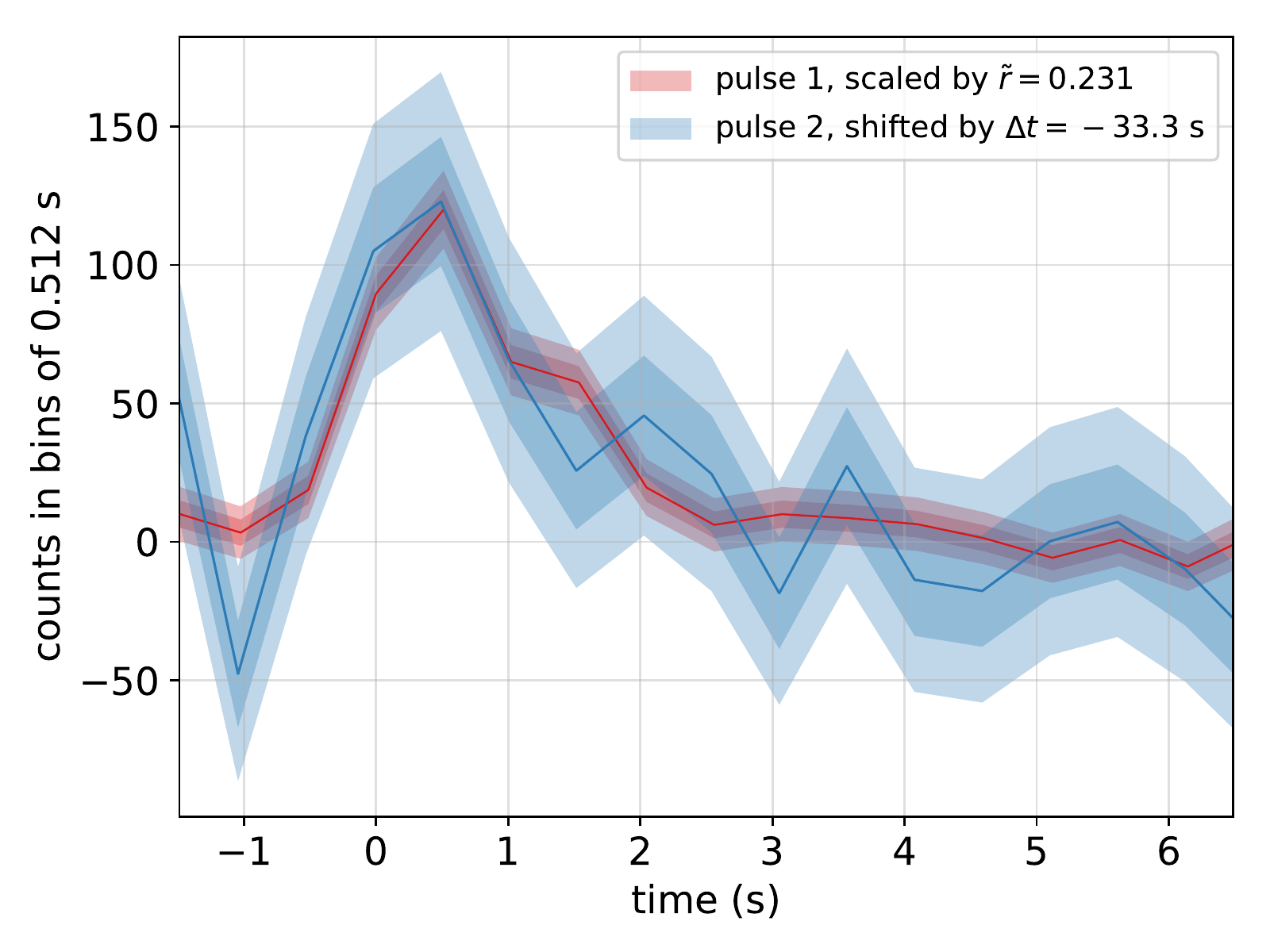}
	\includegraphics[width=0.49\columnwidth]{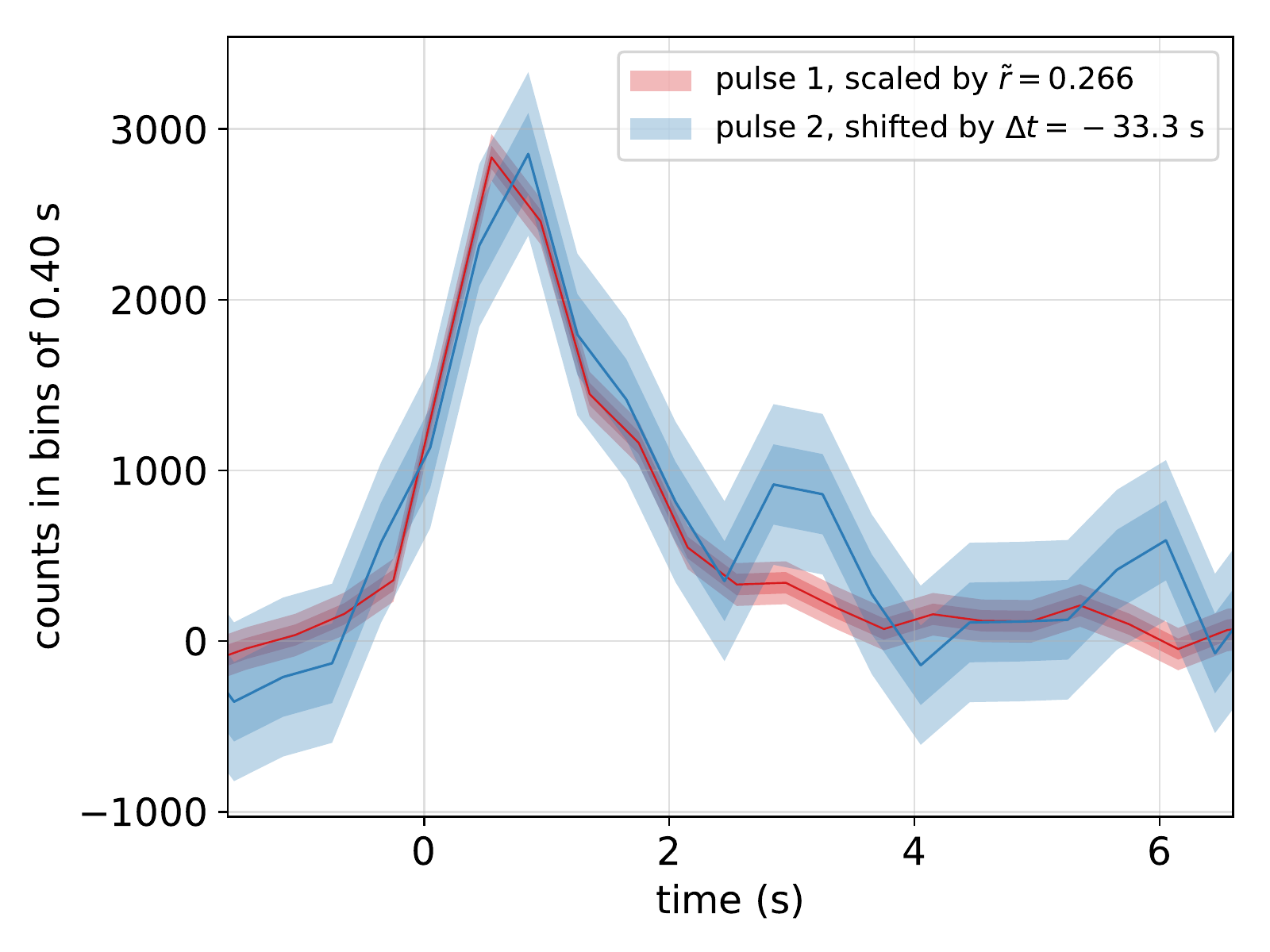}
    \caption{Comparison of the lightcurves of the two pulses. The stronger first pulse is scaled down to match the second pulse. The fainter second pulse is shifted in time to match the first pulse. Dark and light regions are 1 and 2 sigma regions respectively. The figures represent: top-left: NaI (45-300 keV) + BGO (120-400 keV), top-right: NaI (22-800 keV), bottom left: ACS ($>$ 70 keV), bottom-right: BGO (120-400 keV).}
    \label{fig:chi2}
\end{figure}

\subsection{Temporal properties}
\grb consists of two pulses, separated by a quiescent period of about 30 s (Figure \ref{fig:lcfit}). Using the GBM \texttt{targeted search}, we found no emission between the two pulses. For background estimates, we fit a third degree polynomial based on quiescent segments of the lightcurve before, after and in-between  the two pulses.  We also report no discernible emission around 33 s after the second pulse, indicating that this is not a periodic source. We further note that low level emission discernible in the summed lightcurve just before the start of the second pulse (T0+26.5 s to T0+29.6 s) is inconsistent with coming from the location of \grb, and thus it is unrelated.

\grb has a duration of $T_{90}=39.9\pm 3.6 \s$ (10-1000 keV; $T_{90}$ marks the time interval between 5\% and 95\% of the cumulative flux)\footnote{\url{https://gcn.gsfc.nasa.gov/gcn3/30633.gcn3}}. Analyzed separately, the two pulses have consistent duration: the duration of the first pulse is $T_{90,1}=5.31\pm 0.68 \s$ while the second pulse is $ T_{90,2}=3.84\pm 1.64 \s$ long.  Taking the first pulse as a separate GRB, we classify it based on the GBM $T_{90}$ distribution \citep{Bhat+16cat,vonKienlin+20GBM10yrcat}, as a likely long GRB originating from the collapse of a massive star, with a probability of 87\%. Conversely, the likelihood of \grb being a short GRB from a compact binary merger, based on the $T_{90,1}$ information, the observed distribution of \fermi-GBM GRBs and calculated consistently with \citet{Goldstein+17170817a,Rouco+21180418} is 13 \%.

{\it Autocorrelation function} - We measure the time delay between the two pulses using the autocorrelation function of the lightcurve between 10 and 1000 keV, with 64 ms resolution. 
To accurately determine the autocorrelation peak, we fit a 9th degree polynomial  to the peak region. We estimate the uncertainty of the delay by  adding Poisson noise to the original lightcurve \citep[see e.g.][for a similar approach]{Ukwatta+10lag,Hakkila+18smoke}. After repeating the peak finding procedure, we take the uncertainty as the 1$\sigma$ confidence region of the delays of the modified lightcurves and get:
\begin{equation}
    \Delta t_{\rm ACF} = 33.30^{+0.12}_{-0.11} \s.
\end{equation}
This is the first among the many time delay measurements between the pulses.
We note here that this accuracy is typical of what one would expect for a lensed GRB separated by a longer timescale (macrolensing).

The peak of the auto-correlation curve shows a $3.15\sigma$ excess over the smoothed curve calculated using the Savitzky-Golay filter. This excess satisfies the criteria of \citet{Paynter+21lens} for lensing, which classifies \grb as a lensing candidate for further scrutiny.

{\it Spectral lag -} The lag is a measure of the delay between high and low-energy photons \citep[e.g.][]{Norris+00lag}. Typically, a GRB is harder initially, and the higher energy photons (100-300 keV) arrive earlier than the low energy (25-50 keV) photons. This relation is also used to estimate the redshift based on the empirical correlation between 
lag and luminosity \citep{Norris+00lag}.  For the first pulse we find that the lag is $\tau_{\rm lag,1}=221.6^{+123.4}_{-139.6} \ms$, 
while the second pulse has: $\tau_{\rm lag,2}= -89.2^{+255.9}_{-206.5} \ms$. The error on the second pulse is much larger, because of its weakness, especially in the 25-50 keV range.

\subsection{Spectral Analysis}\label{sec:spec}

\begin{figure}
	\centering
	\includegraphics[width=\columnwidth]{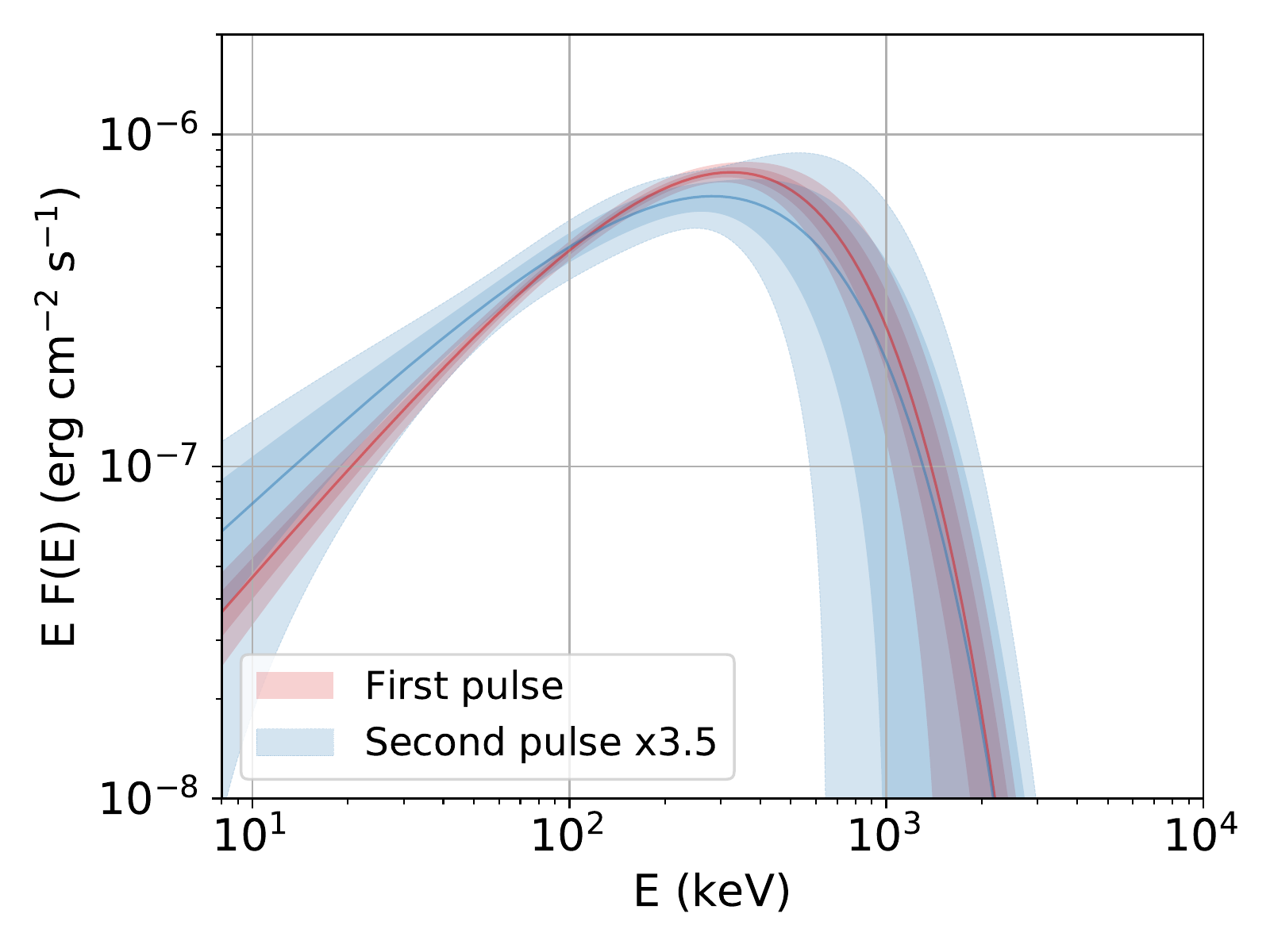}
    \caption{Spectra of the two pulses. The second pulse is shifted to match the first. Shaded regions mark 1 and 2$\sigma$ confidence regions.}
    \label{fig:spec}
\end{figure}

The spectrum of the first pulse is best fit by a power law with exponential cutoff, also known as the Comptonized model (see Table \ref{tab:spec} for the parameters and Figure \ref{fig:spec} for the spectrum). The fluence in the 10-1000 keV range is $F_1=(80.1\pm 0.3)\times 10^{-7} \erg \cm^{-2}$.

The second pulse is weaker, and it is fit equally well by a simple power-law and the Comptonized model. We chose the Comptonized model, because it is more physical (the power law with photon index $>-2$ integrates to infinite energy). The difference in the goodness of fit measure ($\Delta$C-stat) when going from the simpler power law (2 parameters) to the Comptonized model (3 parameters) is $\sim$6. This is just below the $\Delta$C-stat $\approx8$ used in e.g. \citet{Poolakkil+21GBMspcat} to select the more complex model. However, all the parameters of the Comptonized model are well constrained (Table \ref{tab:spec}), which justifies the use of this model. The fluence of pulse 2 is $F_2=(21.6\pm 2.6)\times 10^{-7} \erg \cm^{-2}$.

\begin{figure}
	\centering
	\includegraphics[width=\columnwidth]{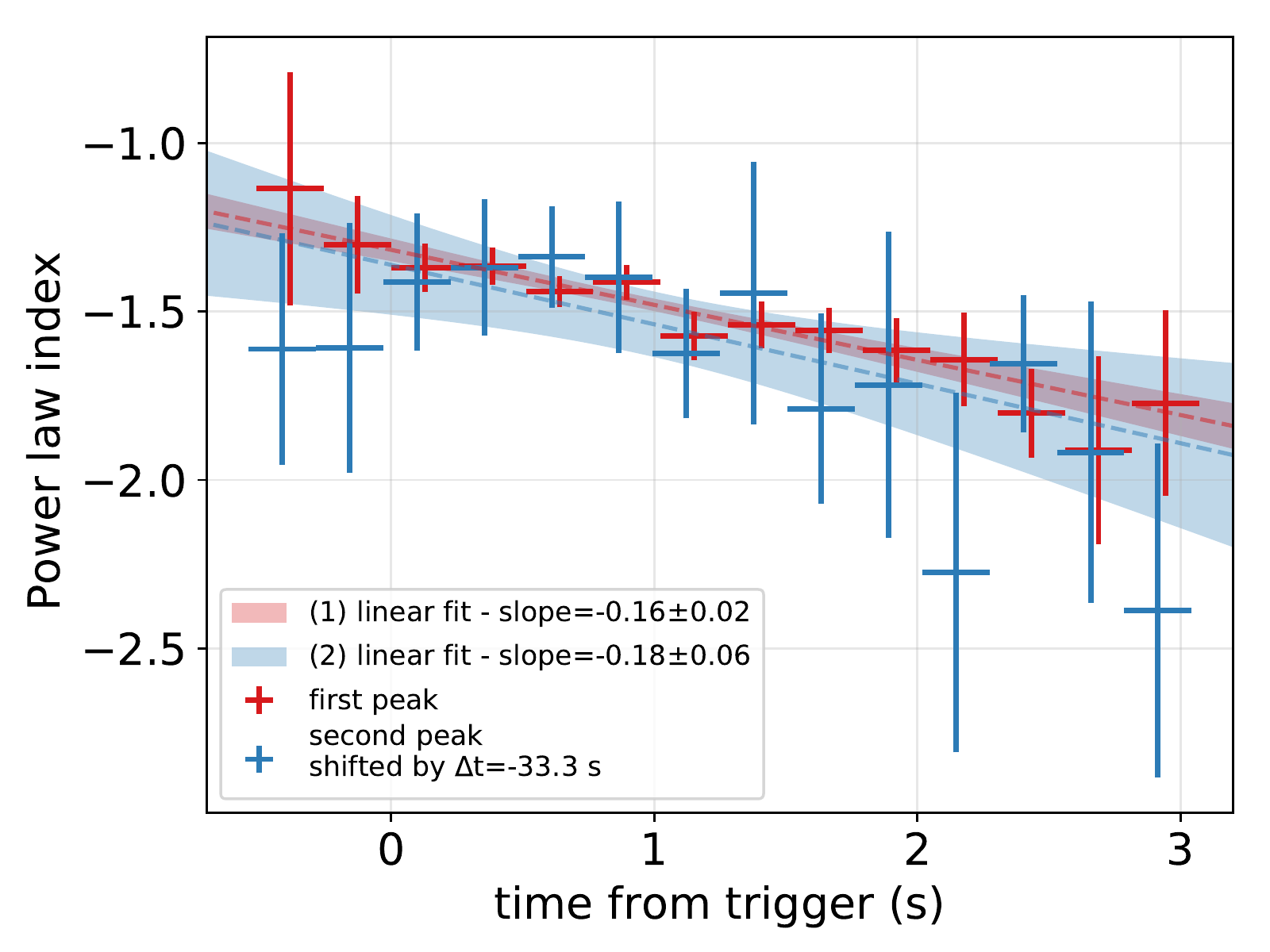}
    \caption{Temporal evolution of the spectral power law index compared for the two pulses. The slopes of the fitted linear functions are indicated in the legend and shaded regions mark 1$\sigma$ confidence intervals.}
    \label{fig:hardness}
\end{figure}

We can also compare the time-resolved spectrum of the two pulses. Because of the relative weakness of the second pulse, we chose to fit the simple power-law model in bins of 0.256 s. We show the temporal evolution of the power-law indices for the two pulses and a linear fit to both as a function of time (Figure \ref{fig:hardness}). We shifted the second pulse by $33.3 \s$ to highlight their similar behavior.

\begin{table*}[]
    \centering
    \begin{tabular}{c|ccccc}
    Interval    & $\Ep$ & photon index & C-stat/dof & photon flux &  Energy flux \\
     (s)   & (keV) &  &  & (ph cm$^{-2}$ s$^{-1}$) &  (10$^{-7}$ erg cm$^{-2}$ s$^{-1}$) \\
    \hline
         -0.256 to 4.096&   $324\pm28$& $-0.88\pm 0.08$ & $373.3/310$ & $10.35\pm0.46$ & $18.4\pm0.7$ \\
         33.024 to 37.376&   $283\pm90$& $-1.10\pm 0.24$ & $295.3/304$ & $3.48\pm0.49$ & $4.96\pm0.59$ \\
    \end{tabular}
    \caption{Spectrum of the two pulses fit by a power law with exponential cutoff.}
    \label{tab:spec}
\end{table*}

\section{Indicators of lensing origin} \label{sec:test}
Proving the gravitational lensing origin involves showing that the two pulses have identical pulse shapes, and their spectra are similar as well. 
It is conceivable that some physical mechanism produces two pulses with identical pulse shapes and spectra with no lensing involved. Without imaging the sources based solely on the gamma-ray observations, a lensing scenario is difficult to prove beyond the shadow of a doubt.

In this section, we will show, however, that the gamma-ray properties of \grb pass all the tests in the literature for a lensing origin. Furthermore, we apply the Bayesian evidence criterion that can select among models, and this method indicates strong evidence in favor of the lensing scenario.

The basic idea for establishing the statistical likelihood of lensing in the temporal domain is comparing two scenarios. First, we assume no lensing. We fit the pulse model and allow every parameter to vary freely. In the alternative scenario, where lensing is assumed, the second pulse is forced to have the same shape as the first one and differ only by a normalization factor and a time delay.

We use two types of pulses \citep{Norris+96pulse,Norris+05pulse}. The first pulse with 5 parameters, referred to as ``N1", is defined as:
 \begin{equation*}
    I_{N1}(t| A, t_{\max},\sigma_r, \sigma_d,\nu) = A \ 
    \left\{ 
  \begin{array}{ c l }
    e^{-\left(\frac{t_{\max}-t}{\sigma_{r}}\right)^\nu} & \textrm{if } t < t_{\rm{max}} \\
    e^{-\left(\frac{t-t_{\max}}{\sigma_{d}}\right)^\nu} & \textrm{if } t > t_{\rm{max}} 
  \end{array}
\right.\label{eq:n1}
\end{equation*}

Here $A$ is the amplitude at peak time, $t_{\max}$.  $\sigma_{r}$ and $\sigma_{d}$ mark the rise and decay timescales of the pulse and $\nu$ is a shape parameter.

The second pulse with 4 parameters, ``N2", has the following temporal dependence: 
\begin{equation}
    I_{N2}(t| A, \xi, \Delta, \tau) = A \ 
    \left\{ 
  \begin{array}{ c l }
    e^{-\xi \left(\frac{t-\Delta}{\tau} +\frac{\tau}{t-\Delta} \right)} & \textrm{if } t > \Delta \\
   0 & \textrm{if } t < \Delta, 
  \end{array}
\right.\label{eq:n2}
\end{equation}
{where} $A$ is the amplitude, $\xi$ is the asymmetry parameter, $\Delta$ is the start time of the pulse, and $\tau$ is a duration parameter.

\subsection{Indirect evidence}
We explore a few properties of \grb that are not direct proofs of lensing, but they are necessary to any such claim.

{\it Hard-to-soft evolution:  } Gamma-ray bursts typically become softer with time \citep[e.g.][]{Kaneko+06batse}. This trend can be observed in the individual pulses of \grb as well (see Figure \ref{fig:hardness}). For GRBs in general, the hard-to-soft evolution can be observed even across pulses.  Specifically, for GRBs that show two pulses separated by a quiescent period, the second pulse is, in general, softer \citep[e.g.][]{Lan+18double,Zhang+12doublegrb}. We find that the second pulse of \grb has the same spectral shape within errors or, subsequently, the same hardness, which is uncommon for typical GRBs.

{\it Leading pulse is brighter: } In case of simple lens models, the light ray traveling closer to the lens arrives later. It has a lower magnification than the first arriving light ray with the larger impact parameter \citep{Krauss+91lens}. We find that the first pulse is indeed visibly brighter  and thus consistent with the expectation from a lensed source. We note that this criterion may not hold for complex lens models \citep{Keeton+09lenstimeorder}.


\subsection{Spectrum of the pulses}
\label{sec:spec}
For the spectral analysis we first selected the interval containing the second pulse by visually identifying contiguous temporal bins with significant signal. %
For the first pulse, we selected a source interval which is the same length as the second pulse (see Table \ref{tab:spec} and Figure \ref{fig:multilc}).
The spectral fits of the two pulses yield consistent spectral shapes within errors: the peak of the energy-per-decade or $\nu F_\nu$ spectrum is $E_{\rm peak,1}= 324\pm 28 \keV $ and $E_{\rm peak,1}= 283\pm 90\keV $. To compare the two spectra, we plot the  1 and 2 $\sigma$ confidence regions of the spectral shapes (Figure \ref{fig:spec}), accounting for the correlations between the parameters. We multiply the second pulse by the fiducial $3.5$ number to show that the two spectral shapes are consistent.

\subsection{Count ratio test}
\label{sec:countratio}
If the pulses are gravitationally lensed, the ratio between pulses should not depend on energy. \citet{Mukherjee+21lens} investigated the count ratio of the two pulses as a function of energy for GRB 950830 and found a $\gtrsim 2\sigma$ inconsistency. 
This test has the advantage that it is independent of the particular spectral shape.

For GBM  we used the 8-channel \texttt{ctime}  data to carry out this test on \grb. We considered all the detectors where the second pulse was visible in any channel (n1, n6 through nb, energy channels 1 through 6, and b0 and b1, energy channels 0 and 1), and data from ACS and BAT (25-350 keV). We find that the ratio of the pulses in all the channels and all three instruments is consistent with the mean value within 1.6 standard deviations (see Figure \ref{fig:hardness}). We thus conclude that \grb passes the count ratio test for lensing.

\subsection{$\chi^2$ test}
\label{sec:chi}
A simple and robust test that doesn't assume any pulse shape was introduced by \citet{Nemiroff+01millilens}. This test was recently applied to the claim of \citet{Kalantari+21lens} on GRB 090717  by \citet{Mukherjee+21lenslctest}. \citet{Mukherjee+21lenslctest} conclude that based on the $\chi^2$ test, the claim of gravitational lensing can be excluded at the $5\sigma$ level.

This test considers the binned lightcurves of the two pulses as representing two distributions and asks if they are consistent with coming from the same parent-distribution. After appropriately re-scaling the first pulse and taking into account the background, we perform a $\chi^2$-test for the hypothesis that the two lightcurves are drawn from the same distribution. 

The test statistic is defined the following way:
\begin{equation}
    \chi^2=\sum_t \frac{(\tilde{r}P_1(t)-P_2(t+\Delta t))^2}{\tilde{r}^2 P_1(t) +P_2(t+\Delta t) + B_1(t)+B_2(t+\Delta t)},\label{eq:chi2}
\end{equation}
where the $t$ is the time, $\tilde{r}< 1$ is the scaling factor, $P_i$ marks the background subtracted counts in the two pulses, {and} $B_i$ {is} the background counts ($i=\{1,2\}$). 

As an example (see Figure \ref{fig:chi2}, top left), we consider data from  $T0-2 \s$ to $T0+7 \s$ interval (first pulse) and compare  it with the interval shifted by $\Delta t =33.3 \s$ (second pulse). We use 0.256 s resolution, summed over the detectors with good signal in the 45-300 keV range, and added the signal of the BGO detectors (120-400 keV). We use the \texttt{tte} instead of \texttt{ctime} data, because the beginning of the first pulse has uneven temporal binning in the \texttt{ctime} data.

First, we find the minimum of the $\chi^2$ expression in Equation \ref{eq:chi2} as a function of $\tilde{r}$ and we get $\tilde{r}=0.231$ (see Figure \ref{fig:chi2}, top-left). Next, we calculate the minimum $\chi^2=24.3$ value for 34 degrees of freedom, which corresponds to a p-value of $0.89$. 
Thus there is no statistically significant difference between the two distributions.

We compare the lightcurves of the two pulses for different temporal resolutions, energy ranges and instruments, in the panels of Figure \ref{fig:chi2}. We consistently find there is no statistically significant difference between the two pulses. For example, \citet{Mukherjee_21210812gcn} performed a preliminary $\chi^2$ analysis on \grb and concluded there is a $\sim 2.8\sigma$ discrepancy between the  two pulses. Using the same energy range, and detector selection  as \citet{Mukherjee_21210812gcn} (assuming in their notation detectors are numbered 1 through 12, and \texttt{ctime} energy channels 1 through 8) we performed the $\chi^2$ test on 512 ms resolution lightcurves and in the energy range 22-800 keV (NaI detectors only, Figure \ref{fig:chi2}, top-right). We find $\tilde{r}=0.231$ minimizes $\chi^2$ at a value of $\chi_2=11.5$ for 16 degrees of freedom (p-value of 0.78), indicating the two lightcurves are consistent with being drawn from the same distribution. The BGO lightcurve (512 ms resolution) yields $\chi^2=14.027 ({\rm dof}=16)$ and p-value of 0.597, and for ACS (0.4 s resolution) $\chi^2=24.662 ({\rm dof}=22)$ and p-value of 0.314 (bottom two panels of Figure \ref{fig:chi2}).

\subsection{Bayesian Model Comparison}
\label{sec:bayes}
Applying an idea from gravitational wave model selection \citet{Paynter+21lens} introduced the Bayesian evidence to compare the lensing scenario to the case where there is no lensing.

In the {\it no-lens} scenario we fit the lightcurve with two pulses, with all the parameters left to vary. We derive the Bayesian evidence $\mathcal{Z_{\rm NL}}$ by integrating the likelihood over the multi-dimensional parameter space. E.g. for the N1 pulse, this involves 10  parameters: $I(t)=I_{N1}(t|A_1,\sigma_{r,1}, \sigma_{d,1}, t_{\max,1},\nu_1)+I_{N1}(t|A_2,\sigma_{r,2}, \sigma_{d,2}, t_{\max,2},\nu_2) $.
In the {\it lensing} scenario the second pulse is constrained. It has  the same shape parameters as the first pulse, only differing in the normalization ($r$) and the shift in the peak time ($\Delta t$), resulting in $5+2$  parameters:
$I(t)=I_{N1}(t|A,\sigma_{r}, \sigma_{d}, t_{\max},\nu)+I_{N1}(t|A/r,\sigma_{r}, \sigma_{d}, t_{\max}+\Delta t,\nu) $. $\mathcal{Z_{\rm L}}$ is the evidence in this case.

Formally, the Bayesian evidence (or simply evidence) has the following meaning \citep[e.g.][]{Speagle20dynesty}: from Bayes' Rule, the probability of the model parameters (in our case the peak times, pulse widths, amplitudes, etc., denoted by $\Theta$), given the observations ($D$) and a model ($M$, e.g. the lensing scenario using the N1 pulse shape) is:  $P(\Theta|D,M)=P(D|\Theta, M) P(\Theta|M)/P(D|M)$. Here $P(D|\Theta, M)$ is the likelihood of the data given the model and its parameters and $P(\Theta|M)$ represents the our prior knowledge of the parameters. The evidence is the denominator in the expression of Bayes' Rule: $\mathcal{Z}=P(D|M)=\int_{V_\Theta}P(D|\Theta,M) P(\Theta|M)d\Theta$. The integral is performed over the hyper-volume $V_\Theta$ constructed from all the parameters.

Calculating the evidence $\mathcal{Z}$ is computationally intensive. It requires integrating the likelihood over a multi-dimensional parameter space. 
We use the \texttt{bilby}  python package \citep{Ashton+19bilby} and \texttt{dynesty} nested sampler \citep{Speagle20dynesty} to carry out the integration over the parameters to find  $\mathcal{Z}$. As in \citet{Paynter+21lens} the difference in $\ln \mathcal{Z}$ is the natural logarithm of the Bayes Factor ($\ln {\rm BF} =\ln\mathcal{Z_{\rm L}}-\ln\mathcal{Z_{\rm NL}}$)  and it can be used to decide between the models. The Bayes Factor is additive, values from different, independent (e.g. different energy ranges) measurements can be added to perform model comparison.

We present the results of this analysis in Table \ref{tab:Zs}. Conveniently, the nested sampling yields also the best fitting flux ratio and time delay.  Data from every instrument and energy range where the second pulse was detectable provided positive Bayesian evidence in favor of the lensing scenario.

\begin{table*}[]
    \centering
    \begin{tabular}{c|ccccc}
    detector & energy range    & pulse & $\ln$(BF)  & r & $\Delta$t  \\
     & (keV)    & model &   &   &  (s) \\
    \hline
    NaI & 8-45  & N1   & $0.96$ & $4.14^{+3.29}_{-1.79}$ & $33.85^{+2.42}_{-2.66}$ \\ 
    NaI  &45-95  & N1   & $0.50$ &$5.53^{+2.67}_{-1.93}$ &$33.37^{+1.05}_{-0.76}$ \\ 
    NaI & 95-300  & N1   & $1.79$ &$5.59^{+2.03}_{-1.27}$ &$33.26\pm0.24$ \\ 
    NaI & 300-500 & N1   & $0.28$ &$5.96^{+2.73}_{-2.81} $ &$33.68^{+3.15}_{-2.39}$ \\ 
    BGO & 120-400 & N1   & $2.29$ &$4.69^{+1.91}_{-1.14}$  &$33.16^{+0.25}_{-0.23}$  \\
    ACS   & $>$70 & N1  &  $2.89$ & $4.26^{+1.71}_{-0.97}$  & $33.37^{+0.23}_{-0.24}$  \\ 
    Swift & 25-350 & N1  &  $0.50$ &$6.02^{+2.71}_{-2.67}$  & $33.12^{+3.55}_{-2.65}$  \\ 
   \hline
    NaI & 8-45  & N2   & $1.94$ & $3.61^{+3.23}_{-1.60}$ & $33.25^{+1.07}_{-1.30}$ \\ 
    NaI & 45-95  & N2   & $2.83$ & $5.09^{+2.48}_{-1.67}$ & $33.11^{+0.79}_{-0.56}$ \\ 
    NaI & 95-300  & N2   & $5.13$ & $5.35^{+1.91}_{-1.24}$ & $33.30^{+0.23}_{-0.25}$ \\ 
    NaI & 300-500  & N2   & $0.81$ & $5.51^{+2.97}_{-2.82}$ & $33.11^{+1.11}_{-1.30}$ \\ 
    BGO & 120-400   & N2   & $5.25 $ &$4.53^{+1.80}_{-1.08}$ &$33.13{\pm0.23}$ \\ 
    ACS &  $>$70  & N2  & $5.18 $   &$4.20^{+1.69}_{-0.93}$ & $33.34\pm0.24$  \\
    Swift & 25-350   & N2  &  $0.55$ &$5.72^{+2.90}_{-2.71}$ & $32.81^{+1.26}_{-1.08}$ \\
    \end{tabular}
    \caption{Bayes factor compilation for different instruments, energy ranges and pulse models using the \texttt{bilby} nested sampling method. The flux ratio and the time delay resulting from the fits are shown in the last two columns. }
    \label{tab:Zs}
\end{table*}

%
\begin{figure}
	\centering
	\includegraphics[width=\columnwidth]{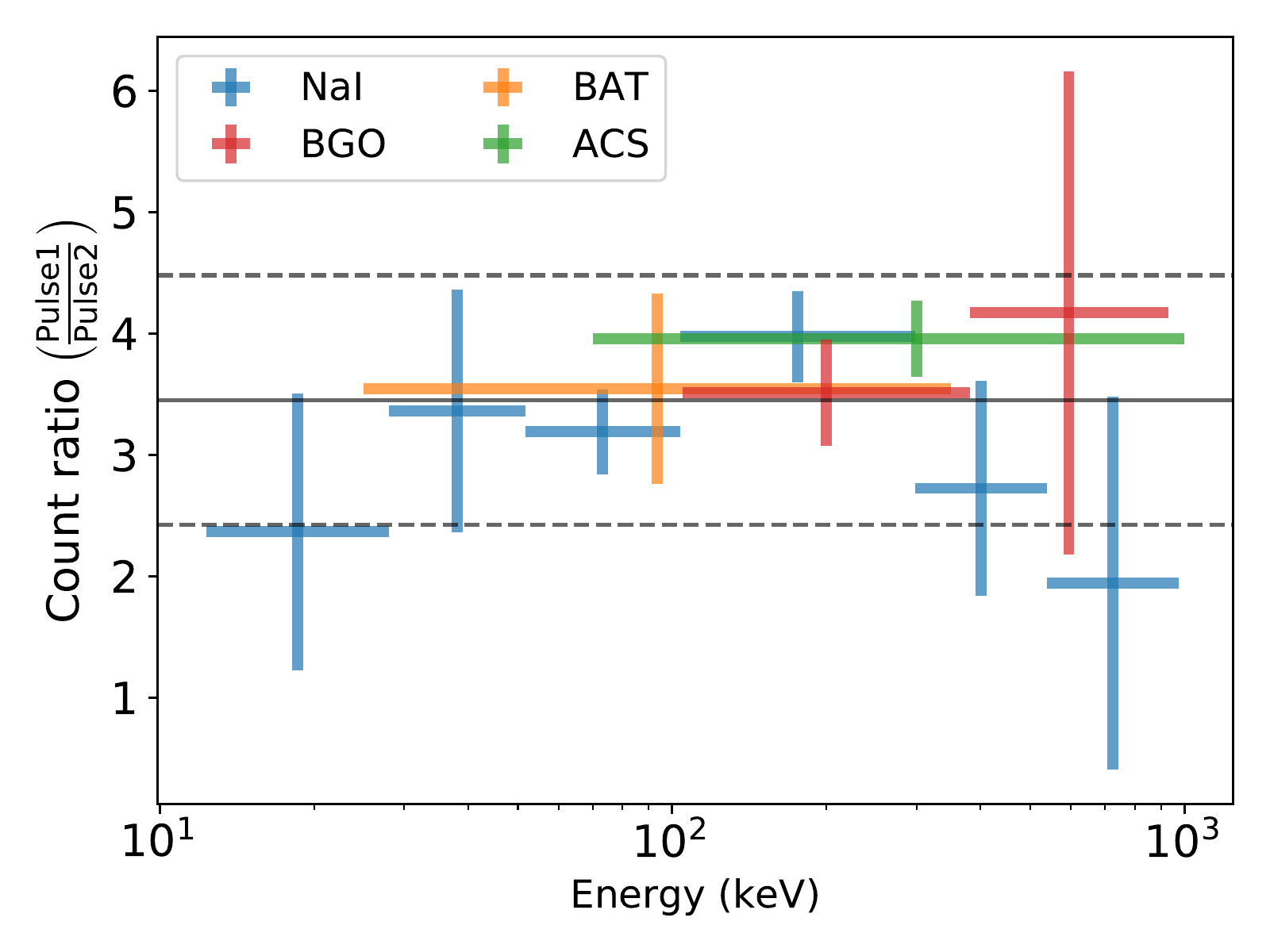}
    \caption{Ratio of counts in the first and second pulse as a function of energy and different instruments. The average count ratio is $3.45\pm1.03$ (horizontal lines.)}
    \label{fig:frtest}
\end{figure}
%
%

%
\begin{figure}
	\centering
	\includegraphics[width=\columnwidth]{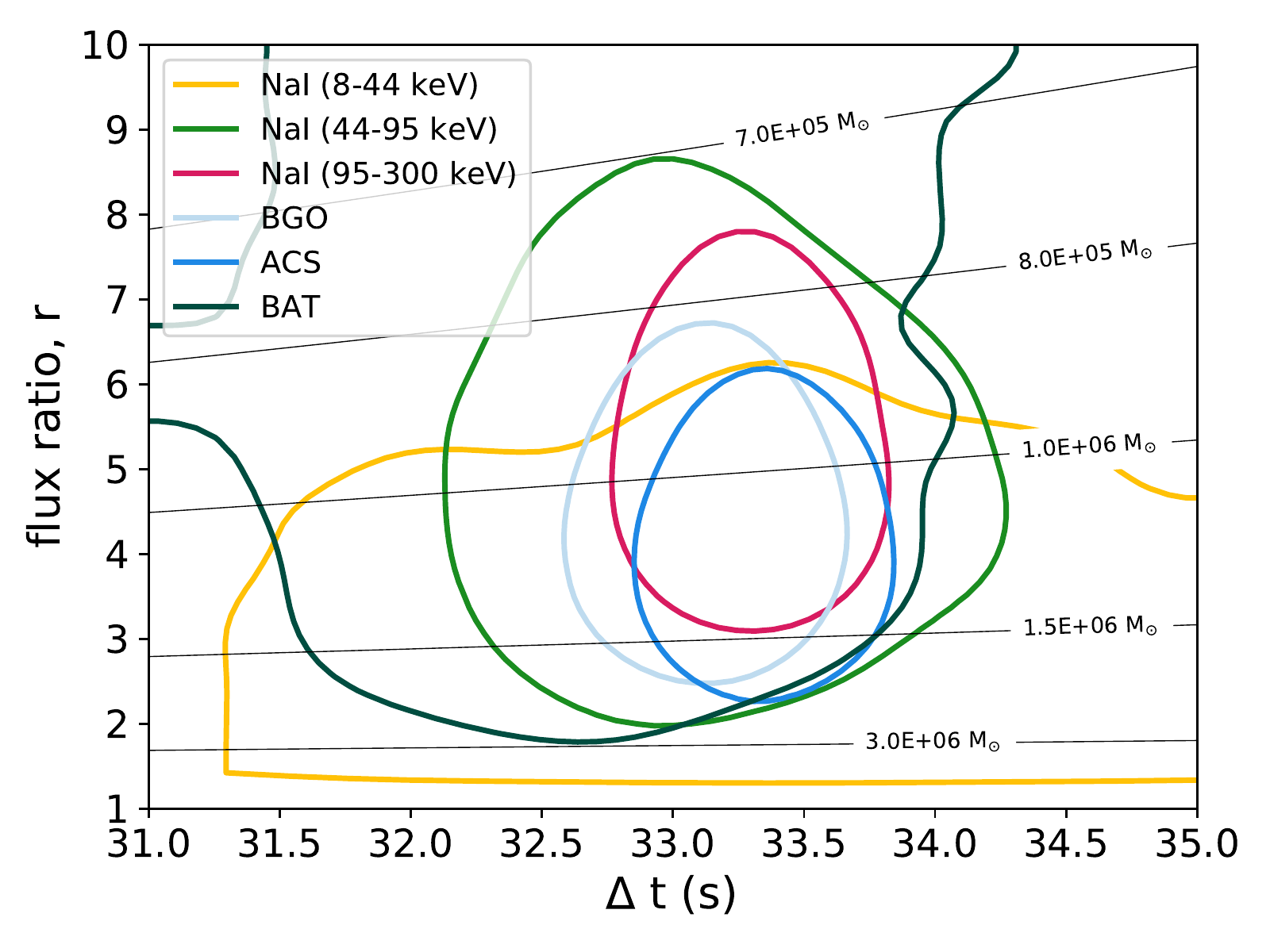}
    \caption{1$\sigma$ confidence region of the time delays and the flux ratios using the N2 pulse model and \texttt{bilby} nested sampling across energy ranges and instruments. 
    Gray contour lines mark the associated $(1+z_l) M_l$ lens mass values. Note the region encompassing $\Delta t\approx33.3 \s$ and $r=4~\rm{ to }~5$ is consistently part of the solutions.}
    \label{fig:dt_r}
\end{figure}
%

%
\begin{figure*}
	\centering
	\includegraphics[width=0.8\textwidth]{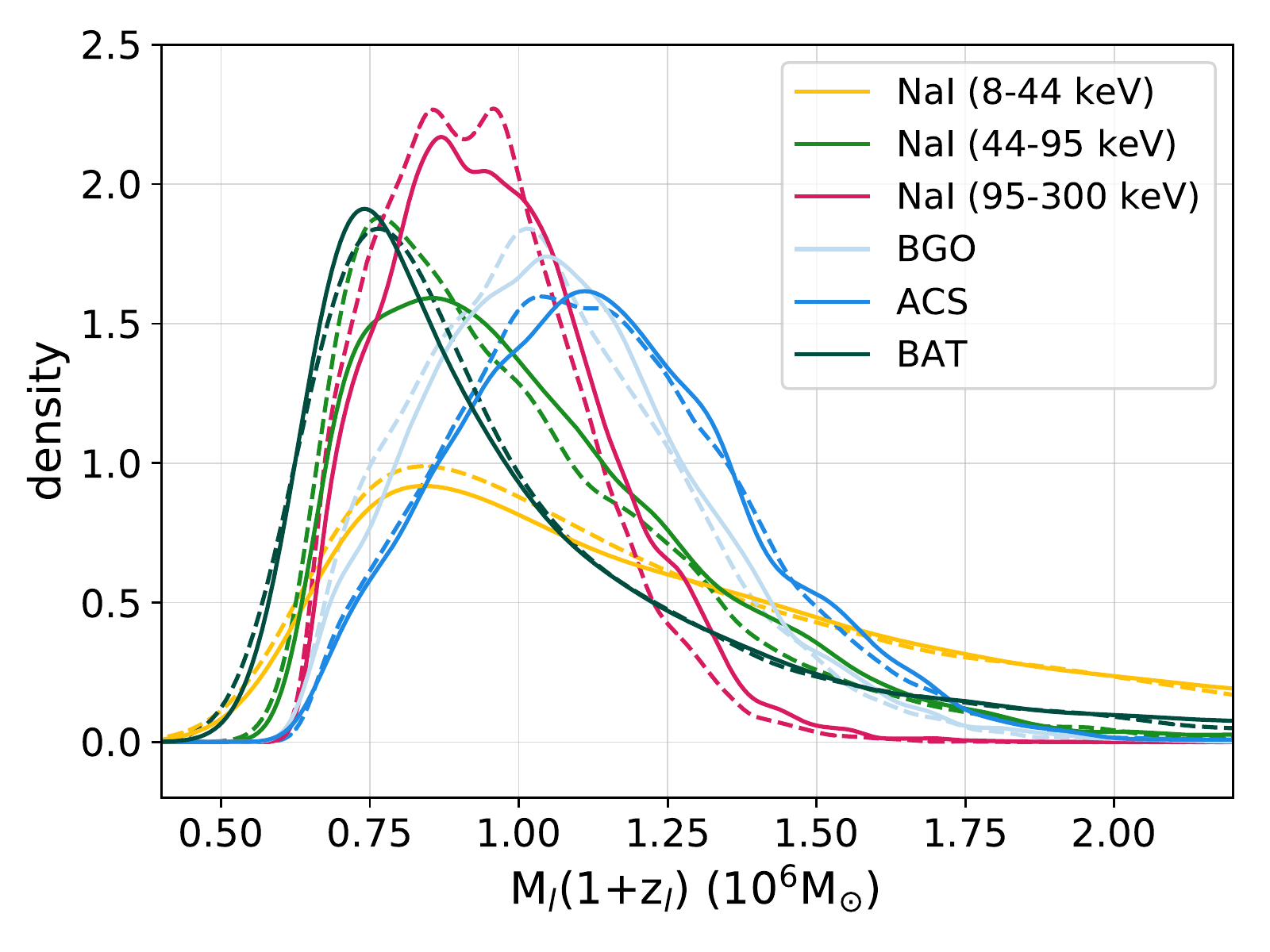}
    \caption{Mass posterior density distribution from the \texttt{bilby} modeling assuming point mass lens. Different colors show different energy ranges or instruments indicated in the legend. Continuous lines represent the N2 pulse model, dashed curves represent the N1 model.}
    \label{fig:masses}
\end{figure*}
%
%

\section{Gravitational lens modeling}

\subsection{Point mass lens}
The simplest mass model is the point mass lens, when the mass of the lens is concentrated in a projected region smaller than the Einstein radius of the source. In this scenario, we can derive the lens mass from the flux ratio and the time delay \citep[e.g.][]{Mao92lens}:

\begin{equation}
   (1+z_l)M_l  = \frac{c^3 \Delta t}{2G}\left(\frac{r-1}{\sqrt{r}} + \ln r
    \right)^{-1}\,,
    \label{eq:dt}
\end{equation}
where $z_l$ is the lens redshift and $M_l$ is the lens mass, $c$ is the speed of light, $G$ is the gravitational constant.

\subsubsection{MCMC lightcurve fitting}\label{sec:mcmc}
We fit the summed NaI and BGO lightcurves with the Markov-chain Monte Carlo method using the \texttt{emcee} \citep{Foreman+13emcee} {python} package with both the N1 and N2 pulses (see Figure \ref{fig:lcfit}). This method cannot select between the lensing and no-lensing scenarios; however, it is fast and robust compared to the more computation-intensive nested sampling (see section \ref{sec:bayes}). We can take the result of the lightcurve fits in the lensing scenario and derive the lens mass, $M_l(1+z_l)$ in the point-mass approximation.

The N1 pulse model leads to a flux ratio of $r=4.47^{+1.06}_{-0.73}$ and delay time of $\Delta t_{\rm l}=33.16^{+0.19}_{-0.21} \s $. The corresponding point lens mass value is:
\begin{equation}
(1+z_l) M_{\rm l}=1.07^{+0.16}_{-0.15} \times 10^6 M_{\odot}\,. 
\end{equation}

For the N2 pulse the flux ratio is $r=4.19^{+0.30}_{-0.25}  $ and the time delay
$\Delta t_{\rm l}=33.11\pm0.06 \s$. These lead to a point mass lens of mass:
\begin{equation}
    (1+z_l) M_{\rm l}=1.13\pm0.06 \times 10^6 M_{\odot}\,. 
\end{equation}

\subsection{Singular Isothermal Sphere (SIS) lens model}
This lens model is characterized by the line of sight velocity dispersion, $\sigma_v$ of its mass distribution. While in the point mass lens case, we could constrain the lens mass, here, it is only possible to restrict the velocity dispersion up to a distance scale $D$,

{\small
\begin{equation}
    \sigma_v=c\left(\frac{1}{32 \pi^2} \frac{c \Delta t}{D (1+z)} \frac{r+1}{r-1} \right)^{1/4}\approx 15 \left(\frac{D}{0.6 \Gpc}\right)^{-1/4} \km \s^{-1},
\end{equation}
}
where $D=D_{OL} D_{LS}/D_{OS}$ and $D_{ij}$ mark the angular diameter distance combinations between the observer (O), lens (L) and source (S), respectively. We assumed a GRB redshift $z_s=1$ and a lens redshift $z_l=0.4$.

A simple mass estimate based on the virial theorem yields a mass $M\approx8\times 10^5~ (\sigma_v/15 \km \s^{-1})^2 (R/10 \pc) M_{\odot}$, where $R=10 \pc$ is an assumed size considered typical for e.g. globular clusters for which the SIS model is a good approximation. The mass is broadly consistent with the point mass lens approximation.

\section{Discussion}\label{sec:discussion}
In the previous section, we performed tests to confirm \grb is affected by strong gravitational lensing. Here we discuss the strengths of each test, analyze the unlensed GRB properties and present future detection prospects.

\subsection{Spectrum}The most basic spectral test for a lensing scenario is the flux ratio test. Because gravitational lensing is achromatic, the flux ratio of the two pulses has to remain constant across energy ranges and different instruments. This is a robust measure because it doesn't depend on the assumed spectrum. 
The only caveat to consider if the GBM detectors' pointing has changed between the two pulses. In that case, the spectral responses change significantly between the pulses and the recorded counts cannot be compared across the emission episodes of \grb. The pointing of the detectors however has not changed by more than 5 degrees between the pulses, which means the detectors' response in the direction of \grb is essentially the same throughout the duration of the burst. The flux ratios across the energy ranges and instruments are clearly consistent with being equal (Figure \ref{fig:frtest}). The weighted mean is 3.45, and we measure the most significant deviation for the ACS data point (green), which is 1.6 standard deviations away.

\citet{Mukherjee_21210812gcn} showed a preliminary analysis of the hardness ratio (HR) for \texttt{ctime} channels 3 and 4 (4 and 5 in their notation) and claimed a 2.2 $\sigma$ discrepancy between the HR of the two pulses. Formally, the count ratio (CR) test is equivalent to the HR test: the hardness ratio of Pulse 1 is HR(P1)=C(P1, Ch2)/C(P1, Ch1), where C(P1, Ch2) denotes the count rate in pulse 1 for channel 2 (higher energy channel for the hardness), and C(P1, Ch1), HR(P2) can be calculated analogously. The count ratio in the energy channel 1 is CR(Ch1) = C(P1,Ch1)/C(P2,Ch1). Thus from CR(Ch1)=CR(Ch2), it follows that HR(P1)=HR(P2). Because there are no strong outliers in the count ratio test, we expect the data to reflect this in the HR test. For the first pulse we find: HR$=1.405 \pm 0.063$ and for the second pulse: HR$=1.127 \pm 0.155$. We confirm the finding of \citet{Mukherjee_21210812gcn} that the second pulse indeed shows a lower HR. However, taking their difference and adding the errors in quadrature, we find that the discrepancy is only $1.66\sigma$, which does not invalidate the lensing scenario.

Next, we carried out a spectral analysis of the two pulses using the GBM data. In the Swift-BAT data, the second pulse was only visible in the summed lightcurve, and INTEGRAL SPI-ACS had only data in 1 energy channel. Therefore we only used \fermi-GBM data for the detailed spectral analysis. The spectral parameters are consistent within errors (Table \ref{tab:spec}), and a plot of the spectral shapes (Figure \ref{fig:spec}) also shows that the two spectra overlap when considering the confidence regions. We thus conclude that the spectra of the pulses are consistent with the lensing interpretation. We note the advantage of the continuous 128 energy channel data of GBM over the 4 channel \texttt{tte} data available for BATSE, for which precise spectral fits were not feasible \citep{Paynter+21lens}.

\subsection{Time history} Gamma-ray instruments have better temporal than spectral resolution. E.g., the number of spectral resolution elements in 10-1000 keV is $\lesssim$10, while the 5 s duration pulse with $\sim$0.1 s resolution has 50 temporal resolution elements, where 0.1 s is the approximate timescale of variations for \grb \citep{Bhat+12mvt}. For this reason, the temporal study of the lensing scenario can provide more constraints. 

The weaker second pulse is closer to the noise level of the detectors. Thus the shorter duration of the second pulse is in line with expectations for a weaker bursts with similar pulse shape. Nonetheless, the duration of the two pulses is still consistent within errors as expected from a lensing scenario.

Independent of any pulse models, we first performed a $\chi^2$  test to compare the two lightcurves. The $\chi^2$ test determines if the two lightcurves are consistent with being drawn from the same distribution.  The $\chi^2$ test showed that there is no significant difference between the lightcurves in different energy ranges and across instruments (Figure \ref{fig:chi2}). We note, however that weakness of the second pulse results in relatively large Poisson errors, and fine temporal structures in the second pulse, if there are any, are washed out. This somewhat reduces the power of this test, showing only that the general shapes of the pulses are consistent.


Next, we introduced pulse models from the literature and fit the lightcurve in different energy bands and different instruments. Using nested sampling, we evaluated the evidence in favor of the lensing scenario using the \texttt{bilby} code. Independent of the detector, energy range, or pulse model  (Table \ref{tab:Zs}), the evidence is consistently {\it for} the lensing scenario as opposed to the non-lensing scenario (BF$>$0 in all cases). We note that the logarithm of the Bayes Factor is not necessarily positive for the model with less parameters (the lensing model in our case). Indeed, e.g. \citet{Wang+21lens} (their Table 1) shows some cases with negative ln(BF).

The Bayes Factor differs depending on which pulse model we apply. We find that in energy ranges where the second pulse is relatively weak, the evidence for lensing is not as strong (but it still favors the lensing). The evidence using the N2 pulse model is more compelling than in the case of the N1 shape. This can be due to the larger number of  parameters in the case of the N1 pulse shape. 

We consider the NaI data alone, and the N2 pulse shape fits. The sum of ln (Bayes Factors) for the NaI detectors yields $\ln({\rm BF})\approx 10.7$. Following \citet{kass+95bayes} and \citet{Thrane+19bayes} we can assign colloquial meaning to this number. A difference of more than 8 is considered {\it strong evidence in favor of} the lensing model.  We conclude that the Bayesian evidence thus supports the lensing interpretation.  We note that the BGO and ACS lightcurves also provide additional evidence and, to a lesser extent, the Swift-BAT data as well.

The nested sampling provides both a selection criterion between the models through the Bayesian evidence and, at the same time, provides the parameters for time delay and flux ratio. We show the values in Table \ref{tab:Zs} and Figure \ref{fig:dt_r}. Different instruments, detectors, and energy ranges all yield a consistent solution, pointing to the lensing origin. Most importantly, we can derive the lens mass for both pulse shapes and arrive at a consistent picture indicating a $\sim 10^6 M_\odot$ lens (Figure \ref{fig:masses}).
The MCMC method yields smaller errors than the nested samplig. This is due to the different fitting approaches of the two methods \citep[see e.g. ][for details]{Speagle20dynesty}.

\subsection{GRB properties}
If \grb had not been lensed, its fluence would have been $F_0=F_1 \left(1-\frac{1}{r}\right)\approx6.2\times 10^{-6}\erg\cm^{-2}$, where $F_1$ is the fluence of the first pulse (see Section \ref{sec:spec}) and we took $r=4.5$ for the numerical value. Using this flux value, we can get a broad range for the possible redshift of this GRB using empirical correlations between gamma-ray properties.  

We scan the 0.1 to 5 redshift range and find that $z\gtrsim 0.5$ is consistent to within one sigma with the Amati relation \citep{Amati2002} between the isotropic equivalent energy, $E_{\rm iso}$ and the redshift-corrected peak energy, $(1+z)\Ep$. While this is a very crude estimate, it is in line with the average measured redshift of GRBs \citep{Bagoly+06redshift,Jakobsson+06redshift} and it is consistent with our fiducial value of $z_s=1$.

The lag-luminosity relationship \citep{Norris+00lag,Gehrels+06strangegrb} provides another estimate of the redshift ($L\propto (\tau_{\rm lag}/(1+z))^{-0.74}$). Taking the median value of the lag (221.6 ms) and scanning the 0.1 to 5 redshift range, we find the redshift of \grb  within the range $0.9<z<1.5$ is consistent at 1$\sigma$ level with the lag-luminosity relation. While this is similarly an empirical relation, it further reinforces that a redshift of $z_s=1$ is a reasonable approximation.

\subsection{Future events} Even though \grb had no multiwavelength follow-up, as more lensed GRB candidates are observed, we expect to have a well-localized counterpart eventually. Identifying the afterglow of a lensed GRB, showing two consistently fading images, would be the smoking gun evidence for the lensing scenario.
The angular separation of the two lensed images on the sky are on the order of the Einstein radius, which for a $10^6 M_\odot$ point mass is $\theta_E=((4GM/c^2) (D_{LS}/D_{OL} D_{OS}) )^{1/2}\approx 3 ~{\rm mas} (M/10^6 M_\odot)^{1/2} (D/0.6 \Gpc)^{-1/2} $ (assuming $z_l=0.4$ and $z_s=1$, mas = "milli-arcsecond"). This falls just short of the resolution of 10m class optical telescopes ($\approx$40 mas). The only conceivable way of resolving the two images is through very large baseline interferometry (VLBI)  radio imaging \citep{Casadio+21lensvlbi}. The sensitivity of VLBI can be as low as $\gtrsim 10~\uJy$ \citep{Venturi+20evn} and radio afterglows are detected in significant numbers at or above this flux \citep{Chandra+12radioag}. VLBI imaging of the two sources will provide additional information on the source and lens redshift and help constrain the lens model. Capturing a lensed GRB with a well-understood lens will allow fully exploit the accurate, sub-second time delay measurement achievable for GRBs and will allow for time-delay cosmography.

\subsection{Lensing object}
It is difficult to identify the type of lens that produced the two pulses of \grb. Possible objects include black holes or globular clusters. Populations of objects can be ruled out based on their number density and their contribution to the total lensing probability \citep{Nemiroff89lensprob}. The total number of millilensed GRBs can be estimated based on a search for lensing candidates in the entire \fermi-GBM GRB catalog. Our goal in this paper was to report only on \grb, and we leave population-level studies for a future work. Nonetheless, we can make a few general observations, based on e.g. the previously mentioned claims by \citet{Kalantari+21lens} and \citet{Wang+21lens,Yang+21lens}. For 1-3 lens events and the total number of \fermi-GBM GRBs, $N>3100$, the lensing rate is $(3-9)\times 10^{-4}$. This is on par with the rate based on BATSE observations by \citet{Paynter+21lens}.

A black hole mass of $M\approx10^6 \Msun$ lies at the lower end of the supermassive black hole population with measured masses \citep[e.g.][]{Woo+10smbhmass} and at the upper end of the intermediate-mass black hole population \citep{Greene+20imbh}. Without detailed counterpart observations it is unclear in which group, if any, the lens of \grb belongs to. Further lensed events however can provide essential constraints on the rates and origin of black holes in this mass range.

Globular clusters are similarly good lens candidates and their masses can indeed reach $10^6 M_{\odot}$ \citep{Baumgardt+18globularcluster}. A SIS model is a good approximation to the velocity dispersion in a globular cluster. \citet{Paynter+21lens} found, however, that even globular clusters with one order of magnitude smaller mass, $\sim 10^5 M_\odot$, do not exist in sufficient numbers to produce the 1/2700 rate of lensed GRBs for BATSE. In our case, we require similar lensing probabilities but with $\sim 10^6 M_\odot$ globular cluster population. $10^6 M_\odot$ globular cluster lies above the approximately $2\times10^5 M_\odot$ turnover mass in the  mass function \citep{Jordan+07globularclusters}. This means that the larger $10^6 M_\odot$ mass cannot compensate in total lensing probability the drop in number density. We thus conclude that the globular cluster lens can be tentatively ruled out, and a point mass lens e.g. a black hole is more likely. For more definitive statements on the nature of the lens precise localization and high resolution observations will be necessary.

\section{Conclusion}\label{sec:conclusion}

In this paper, we presented multiple lines of evidence for \grb being gravitationally lensed. The two peaks in \grb have consistent spectrum, time profile, and spectral evolution. We determined the flux ratio and time delay with multiple methods and arrived at a consistent picture. The first pulse is approximately 4.5 times brighter and the delay between the pulses is $33.3 \s$. Assuming a point mass lens, this flux ratio and delay corresponds to a lens mass of $(1+z_l)M_l=10^6 M_\odot$. There are only a few unchallenged claims in the literature for lensed GRB lightcurves. \grb presents the first strong evidence for lensing a long GRB with a flux ratio larger than 2. Future events will benefit from high resolution radio observations for definitive proof of lensing origin and detailed lens modeling.

{\it Acknowledgments: }
{The authors thank the anonymous referee for a prompt report.}
PV thanks Michael S. Briggs, Zsolt Paragi, S\'andor Frey, Rosa Leticia Becerra for input and discussions and support from NASA grants 80NSSC19K0595 and NNM11AA01A.  NF acknowledges financial  support  from UNAM-DGAPA-PAPIIT  through  grant IN106521.  This work used INTEGRAL public data provided by Dr. V. Savchenko through the INTEGRAL Science Data Centre infrastructure hosted at the University of Geneva.


\bibliographystyle{aasjournal}



\end{document}